\documentclass[showpacs,preprint,amsmath,amssymb,floatfix]{revtex4}

\usepackage{graphicx} 
\usepackage{dcolumn}           
\usepackage{bm}                

\begin{document}
\title{ Unfolding Polyelectrolytes in Trivalent Salt Solutions Using 
DC Electric Fields: A Study by Langevin Dynamics Simulations}  
\author{Yu-Fu Wei} 
\author{Pai-Yi Hsiao} 
\email[corresponding author; e-mail: ]{pyhsiao@ess.nthu.edu.tw}
\affiliation{Department of Engineering and System Science, 
National Tsing Hua University, Hsinchu, Taiwan 300, R.O.C.}
\date{\today} 

\begin{abstract} 
We study the behavior of single linear polyelectrolytes
condensed by trivalent salt under the action of electric fields through
computer simulations. The chain is unfolded when the strength of the electric
field is stronger than a critical value. This critical electric field follows a
scaling law against chain length and the exponent of the scaling law is
$-0.77(1)$, smaller than the theoretical prediction, $-3\nu/2$ [Netz, Phys. Rev.
Lett.  90 (2003) 128104], and the one obtained by simulations in tetravalent 
salt solutions, $-0.453(3)$ [Hsiao and Wu, J. Phys.  Chem. B 112 (2008) 13179]. 
It demonstrates
that the scaling exponent depends sensitively on the salt valence. Hence, it is
easier to unfold chains condensed by multivalent salt of smaller valence.
Moreover, the absolute value of chain electrophoretic mobility increases
drastically when the chain is unfolded in an electric field.  The dependence of
the mobility on electric field and chain length provides a plausible way to
impart chain-length dependence in free-solution electrophoresis via chain
unfolding transition induced by electric fields. Finally, we show that, in
addition to an elongated structure, a condensed chain can be unfolded into an
U-shaped structure. The formation of this structure in our study is purely a
result of the electric polarization, but not of the elasto-hydrodynamics
dominated in sedimentation of polymers. 
\end{abstract}

\maketitle

\section{Introduction} 

To well understand the properties of charged macromolecules in electric
fields, including the conformation and the mobility, is very important in many
domains of researches such as in polymer science, in biophysics, and in
microfluidics, for the reason of a large variety of
applications~\cite{viovy00}.  Applying electric fields stays at the center of
the techniques to manipulate charged macromolecules. It can be also used as
a tool to separate molecules by sizes. However, for the latter case, experiments 
are usually performed in a sieving matrix such as in gel, instead of in a 
solution~\cite{viovy00,kleparnik07,cottet98}.  This is because the free
draining effect in an electrolyte solution produces an electrophoretic mobility
independent of the chain length of macromolecules under a typical electrophoretic 
condition~\cite{olivera64}.  Nevertheless, researchers continue to devote their
efforts in finding ways for size separation in free solutions for the reason of 
its high throughput and applications in microfluidics. 

In 2003, Netz proposed a new strategy to achieve this goal by unfolding
condensed polyelectrolytes (PEs) in electric fields~\cite{netz03a,netz03b}.  
He predicted that the chain mobility increases when a condensed PE chain
unfolds in an electric field, and the critical electric field to unfold a 
chain, $E^*$, depends on the chain length $N$, following the scaling law $E^*
\sim N^{-3\nu/2}$ where $\nu$ is the chain swelling exponent.  
Therefore, longer chains will be unfolded and separated out earlier  
when the applying electric field slowly increases.  
His idea has been recently verified by simulations~\cite{hsiao08} in which PE 
chains were condensed into globules by tetravalent salts and then stretched in 
electric fields. A more general form of the scaling law has been proposed in 
the study, reading as $E^* \sim V^{-1/2}$ where $V$ is the ellipsoidal volume 
calculated from the three eigenvalues of the chain gyration tensor. According 
to the scaling law obtained by the simulations, an electric field of 2kV/cm 
should be applied to unfold collapsed PEs of chain length of order 
$10^6$. This electric field is relatively strong.

For practical reason, we wish $E^*$ to be as small as possible. One way to
reduce $E^*$ is to make less compact the condensed chain structure.  This aim
can be achieved, for example, by increasing temperature, by performing experiments
in high-dielectric solutions, or by using weak condensing agent to collapse 
PEs. In this paper, we choose the last method, using trivalent salt as the condensing 
agent, and study the static and dynamic properties of chains and the unfolding 
electric field.  Since the electrostatic interaction with trivalent salt is $25\%$ 
weaker than that with tetravalent salt, some unexpected situations may take place. 
A key question is to know if $E^*$ still follow the same scaling law of the strong
condensation as shown in Netz's~\cite{netz03a,netz03b} and
in Hsiao and Wu's study~\cite{hsiao08}.  The rest of this paper is organized as follows.  
In Section II, we describe our model and simulation setup. In Section III, we
present our results. The discussed topics include the degree of unfolding, the
critical electric field to unfold a chain, the electrophoretic mobility, the
distribution of condensed trivalent counterions on a chain, and the chain conformation
after unfolding.  We give our conclusions in Section IV.

\section{Model and simulation setup}

Our simulation system contains a single polyelectrolyte and trivalent salt,
placed in a rectangular box with periodic boundary condition.  The
polyelectrolyte dissociates into a polyion chain and many counterions.  The
polyion is modeled by a bead-spring chain, consisting of $N$ beads;  each bead
carries a $-e$ charge where $e$ is the elementary charge unit.  The counterions
are modeled by spheres; each carries $+e$ charge.  The trivalent salt 
dissociates into trivalent cations (counterions) and monovalent 
anions (coions); these
ions are also modeled by charged spheres.  Solvent is treated as a uniform
dielectric medium with dielectric constant equal to $\epsilon_r$.  Three kinds
of interaction are considered: the excluded volume interaction, the Coulomb
interaction, and the bond connectivity.  The excluded volume interaction is
modeled by a purely repulsive Lennard-Jones potential \begin{equation}
U_{ex}(r)=\left\{\begin{array}{ll} 4\varepsilon_{LJ} \left[
\left(\sigma/r\right)^{12}- \left(\sigma/r\right)^6 \right] +\varepsilon_{LJ} &
\mbox{for } r \leq 2^{1/6} \sigma \\ 0 & \mbox{for } r > 2^{1/6} \sigma
\end{array}\right.  \end{equation} where $r$ is the distance between two
particles, $\varepsilon_{LJ}$ is the interaction strength, and $\sigma$ denotes
the diameter of a particle.  We assumed that all the beads and spheres have
identical $\varepsilon_{LJ}$ and $\sigma$.  We set $\varepsilon_{LJ}=k_B T/1.2$
where $k_{B}$ is the Boltzmann constant and $T$ is the temperature.  The
Coulomb interaction is \begin{equation} U_{coul}(r)=\frac{Z_i Z_j \lambda_B k_B
T}{r} \label{eqcoulomb2} \end{equation} where $Z_i$ and $Z_j$ are the valences
of the two charges and $\lambda_B=e^2/(4\pi\epsilon_r \epsilon_0 k_B T)$ is the
Bjerrum length, at which two unit charges have the Coulomb interaction
tantamount to the thermal energy $k_B T$.  We set $\lambda_B$ to be $3\sigma$
to simulate highly charged PEs, such as polystyrene sulfonate.  
$U_{coul}$ was calculated by PPPM Ewald method.
Two adjacent beads (monomers) on the chain are connected by the bond
connectivity,  modeled by a finitely extensible nonlinear elastic potential
\begin{equation} U_{bond}(b)= -\frac{1}{2} k_{b} b_{max}^2 \ln
\left(1-\frac{b^2}{b_{max}^2}\right) \end{equation} where $b$ is the bond
length, $b_{max}$ is the maximum bond extention, and $k_{b}$ is the spring
constant.  We set $b_{max}=2\sigma$ and $k_{b}=5.8333 k_B T/\sigma^2$.  The
average bond length under this setup is about $1.1\sigma$.  An external uniform
electric field $\vec{E}$ is applied, toward $x$ direction.  The equation of
motion of a particle is described by the Langevin equation: \begin{equation} m_i
\ddot{\vec{r}}_i = -\frac{\partial U}{\partial \vec{r}_i} -m_i\gamma_i
\dot{\vec{r}}_i + Z_i e E \hat{x} +\vec{{\eta}}_i(t) \label{eqlangevin}
\end{equation} where $m_i$ is the mass of the particle $i$, $\vec{r}_i$ is its
position vector, $m_i\gamma_i$ is the friction coefficient, and $\vec{\eta}_i$
simulates the random collision by solvent molecules.  $\vec{\eta}_i(t)$ has
zero mean over time and satisfies the fluctuation-dissipation theorem:
\begin{equation} \left< \vec{\eta}_i(t) \cdot \vec{\eta}_j(t') \right> = 6 k_B
T m_i\gamma_i\delta_{ij} \delta(t-t') \label{eqflucdiss} \end{equation} where
$\delta_{ij}$ and $\delta(t-t')$ are the Kronecker and the Dirac delta
function, respectively.  The temperature control is incorporated according to
this theorem.  We assumed that the particles have the same mass $m$ and damping
constant $\gamma$.  We set $\gamma=1 \tau^{-1}$ where $\tau=\sigma
\sqrt{m/(k_BT)}$ is the time unit.  We know that the dynamics of polymers in
dilute solutions is described by Zimm model~\cite{doi86book}.  However,
when an electric field is applied in a typical electrophoretic condition, the
hydrodynamic interaction is largely canceled out due to the opposite motions of
the ions in the electrolyte solution~\cite{viovy00,long96,tanaka2002}.
Therefore, in this study we neglected the hydrodynamic interaction.
Hydrodynamic interaction is important only when the chain length is very
short~\cite{Stellwagen03,grass08}.

We varied the chain length (or the number of monomer) from $24$ to $384$ and
studied the static and dynamic properties of PE under the action of an electric
field, up to a field strength $E=2.0$ $k_BT/(e\sigma)$.  We set the monomer
concentration $C_m= 0.0001 \sigma^{-3}$.  In order to keep $C_m$ constant, the
size of the simulation box needs to  change with $N$.  Instead of using a cubic
simulation box, we chose a rectangular parallelepiped of $1.6N \sigma\times
79.06\sigma \times 79.06\sigma$, where the box size in the field direction is
linearly proportional to $N$ to prevent overlap under periodic boundary
condition when the chain unfolds.  The added salt concentration was fixed
at $C_s=C_m/3$, the equivalence point $C_s^*$. It has been shown that at this salt
concentration, the chain collapsed into a compact globule structure, in the
absence of electric field~\cite{hsiao06a,hsiao06c}, with its effective chain charge
almost being neutralized.  We performed  Langevin dynamics
simulations~\cite{lammps} with integrating time step equal to $\vartriangle
t=0.005 \tau$.  We ran firstly $10^6$ to $10^7$ time steps to bring the system
to a steady state and then ran $10^8$ time steps to cumulate data for analysis. To
simplify the notation, we assign in the following text that $\sigma$, $m$, and 
$k_B T$ are the unit of length, mass, and energy, respectively.  Therefore, 
the concentration will be described in unit of $\sigma^{-3}$, the strength of 
electric field in unit of $k_BT/(e\sigma)$, and so forth.

\section{Results and Discussions} 
\subsection{Degree of unfolding} 

\begin{figure} \centering 
\includegraphics[width=0.6\textwidth,angle=270]{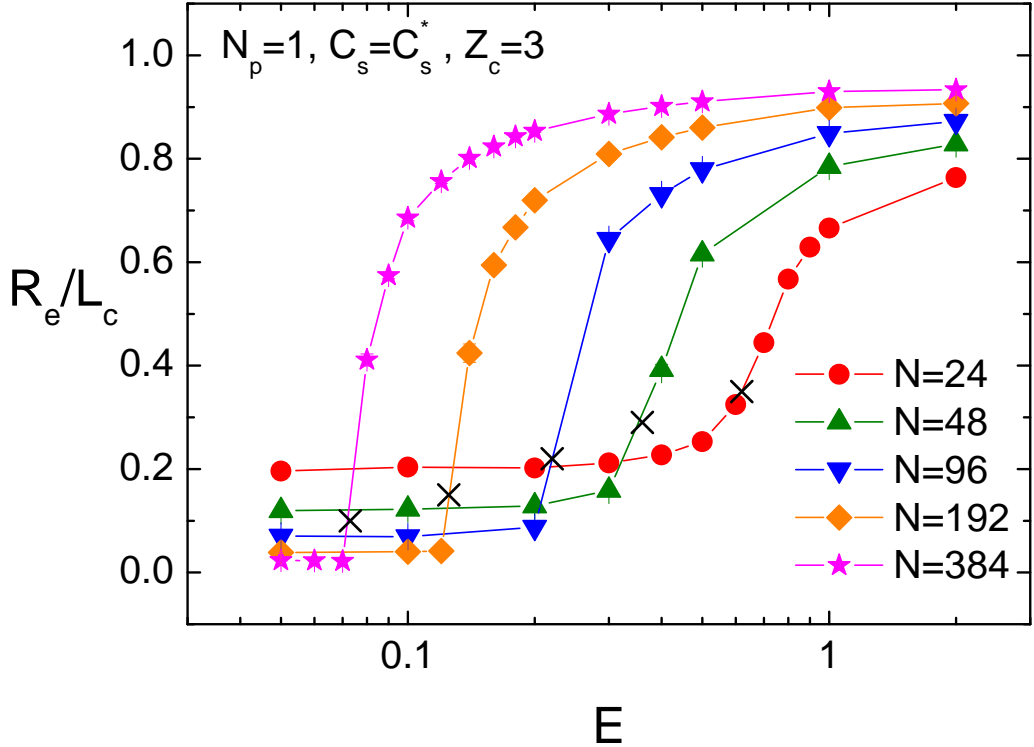}
\caption{$R_e/L_c$ as a function of $E$ at $C_s=C_s^*$ for different chain
length $N$. The symbol `x' denotes the inflection point of curve.} \label{f:ReoLc-E} \end{figure}

We start from studying the chain conformation under the action of an electric
field.  The degree of unfolding, defined as the ratio of the end-to-end
distance $R_e$ of chain over the chain contour length $L_c=(N-1)b$, is used to
characterized the conformation.  The results are plotted in
Fig.~\ref{f:ReoLc-E} as a function of $E$.

Each curve in the plot denotes the variation of $R_e/L_c$  for a given chain
length $N$.  We can see that when the electric field is weak, the ratio is a
constant.  This indicates an unperturbed conformation of chain and
the chain remains in a collapsed structure. An abrupt increase appears when
$E$ is increased over some critical value $E^*$. 
$R_e$ can become as large as $90\%$ of $L_c$ if the applied field is very strong. 
This indicates a structural transition from a collapsed structure to an elongated structure.  
We noticed that the value of $E^*$ depends on the chain length.  The
longer the chain length, the smaller the $E^*$ will be.  Moreover, this
structural transition happens in an interval of $E$.  The size of the interval
decreases with increasing chain length.  Although the transition becomes sharper 
when chain length is long, $R_e$ increases in a continuous way with $E$, 
which suggests a second-order transition.

\subsection{Critical electric field $E^*$} 

\begin{figure} \centering 
\includegraphics[width=0.6\textwidth,angle=270]{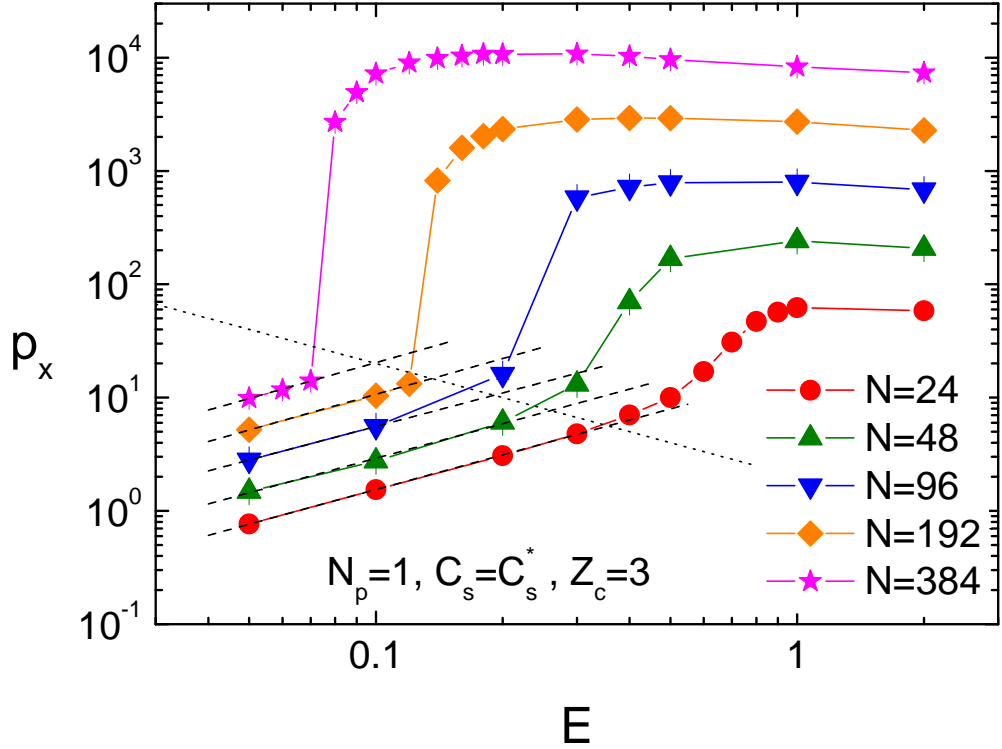}
\caption{$p_x$ as a function of $E$ for different chain length $N$.  The dotted
line denotes the equation $p_xE/2=k_BT$.} \label{f:px-E} \end{figure}

The dependence of the critical electric field $E^*$ on chain length $N$ has
been investigated in salt-free~\cite{netz03a,netz03b} and in
tetravalent salt solutions~\cite{hsiao08}.  Both of these studies showed that
$E^*$ scales roughly as $N^{-0.5}$ to unfold a condensed chain.  It is now important to
know if this scaling law is valid for a PE chain condensed by trivalent
counterions.   To verify it, we follow firstly the method proposed by
Netz~\cite{netz03a,netz03b}:  $E^*$ is calculated by equating the polarization
energy $U_{pol}=\vec{p} \cdot \vec{E}/2$ and the thermal fluctuation energy
$k_B T$.  Here $\vec{p}$ is the dipole moment of the PE-ion complex induced by
the electric field and calculated by $\vec{p} =\sum_i Z_i e
(\vec{r}_i-\vec{r}_{cm})$ where $\vec{r}_i$ is the position vector, running
over all the particles inside the complex, and $\vec{r}_{cm}$ is the center of
mass of the PE.  The complex is considered as a set of particles, including
monomers and ions, inside the region of a worm-shaped tube which is the union
of the jointed spheres of radius $r_t=3$, centered at each monomer center.
The component of $\vec{p}$ at the field direction, $p_x$, is plotted against
the field strength $E$ in Fig.~\ref{f:px-E}.

\begin{figure} \centering 
\includegraphics[width=0.6\textwidth,angle=270]{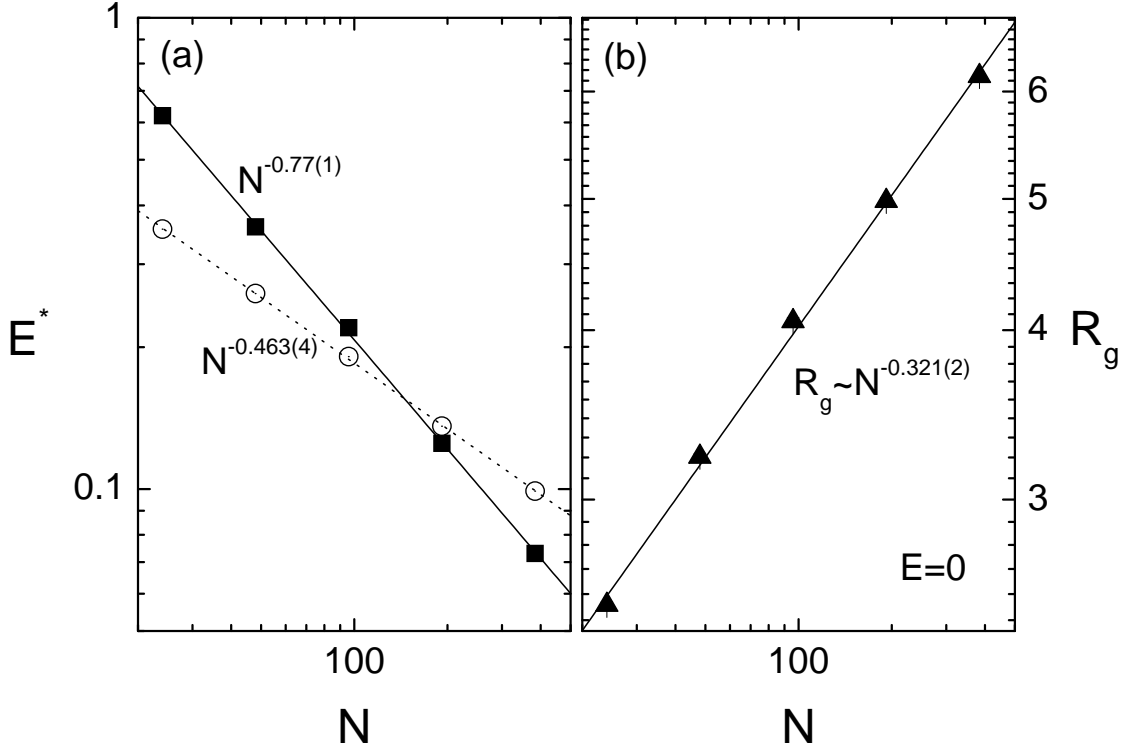}
\caption{(a) $E^*$ vs.~$N$ where the open circles denote the data obtained by
Netz's method and the close squares denote the ones obtained from the
inflection points.  (b) Radius of gyration $R_g$ vs.~$N$ in zero electric field.  } \label{f:scale}
\end{figure}

As seen in the log-log plot, $p_x$ increases linearly with $E$ with a slope 
equal to 1, when $E$ is small.  This is the well-known linear response of a dielectric
object,  $p_x=\alpha E$, which has been reported in the previous
studies~\cite{netz03b,hsiao08}.  But different to the previous, we
found that this linear region terminates before intersecting with the dotted
line which denotes the relation $p_xE/2=k_BT$, specially when the chain length is long.
This is simply because the binding force to condense the PE chain in the
trivalent salt solutions is weaker than in the tetravalent salt~\cite{hsiao08}.
For the system studied by Netz~\cite{netz03a,netz03b}, the chains were strongly
condensed because of the un-realistically strong Coulomb coupling chosen by him. Therefore, his 
method can be used only as a rough estimation of $E^*$ for the case of strong condensation
but not suitable for the weak condensation. If we continue going with
his method and calculate the intersection between the extended linear region and 
the dotted line, we will find that $E^*$ scales as $N^{-0.463(4)}$ 
(see open circles in Fig.~\ref{f:scale}(a)). 
This scaling law seems to follow the prediction of Netz, $N^{-3\nu/2}$,
because the chain swelling exponent in zero electric field is $\nu=0.321(2)$
for this case (cf. Fig.~\ref{f:scale}(b)). Nevertheless, $E^*$ obtained by this
method is actually overestimated, going much over away the linear response region,
specially when the chain is long.

To give a more accurate estimation of $E^*$, we follow here the definition of
the unfolding electric field by taking simply the electric field at the
inflection point of the curve $R_e/L_c$ vs.~$E$.  The inflection
point on each curve in Fig.~\ref{f:ReoLc-E} is indicated by the symbol `x'.  
The scaling law
obtained by this method reads as $E \sim N^{-0.77(1)}$ (see close squares in
Fig.~\ref{f:scale}(a)).  The exponent $-0.77(1)$ is significantly smaller than
the one obtained by the Netz's method.  Therefore, $E^*$ is smaller than 
Netz's prediction for a long chain and it is easier to unfold a PE chain
in trivalent salt solutions.  For example, for a chain of length $10^6$, $E^*$
is $1.76\times 10^{-4}$ according to the scaling law. This $E^*$ corresponds to
about 185 V/cm, much smaller than 2kV/cm predicted for
the chains condensed by tetravalent salt in simulations~\cite{hsiao08}.
Our results show that the valence of the condensing agent plays an important role in
determination of the scaling law.  There must exists more complicated mechanism
to polarize and to unfold a PE chain in an electric field than our thinking.
This mechanism will be investigated in detail in the future.  

\subsection{Electrophoretic mobility and ion condensation}

\begin{figure} \centering 
\includegraphics[width=0.6\textwidth,angle=270]{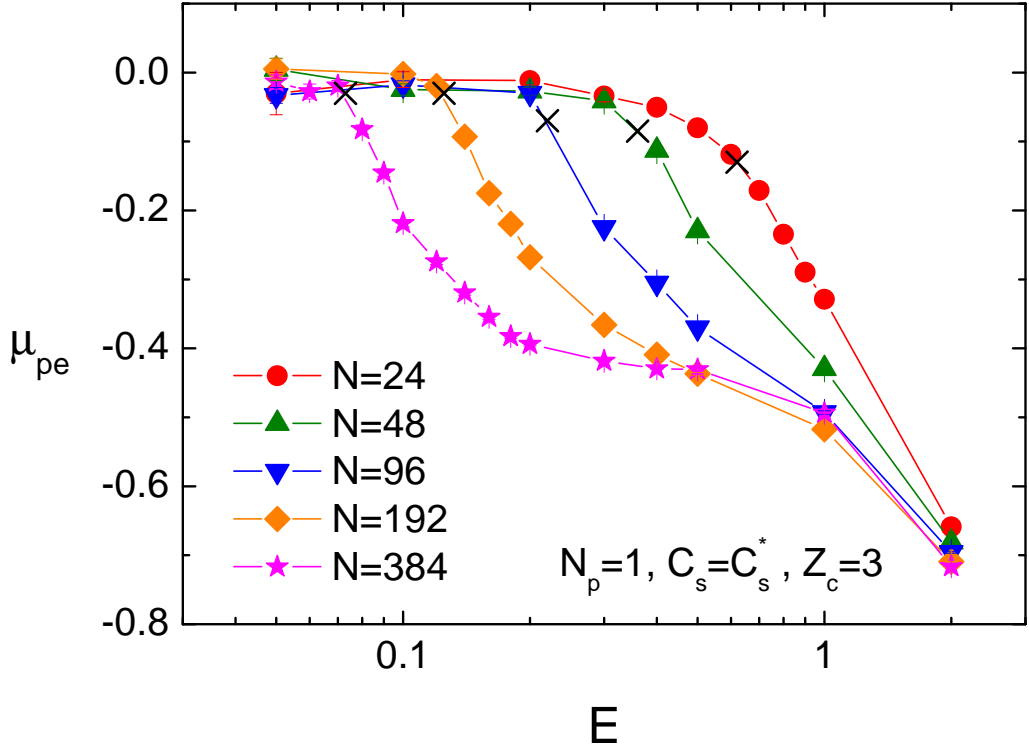}
\caption{$\mu_{pe}$ as a function of $E$ for different chain length $N$.}
\label{f:mu-E} \end{figure}

We now study the electrophoretic mobility $\mu_{pe}$ of PE chain in
electric fields of different strength and show how $\mu_{pe}$ changes with $E$
when the chain is unfolded to an elongated structure.  $\mu_{pe}$ was
calculated by $v_{pe}/E$ where $v_{pe}$ is the velocity of the center of mass
of the chain in the field direction.  The results are shown in
Fig.~\ref{f:mu-E}.

In weak electric fields, $\mu_{pe}$ is nearly zero, indicating that the PE chain is
effectively charge-neutral, as reported in experiments~\cite{bloomfield96}.
While $E$ is increased over $E^*$, $\mu_{pe}$ turns to be negative and the chain
starts to drift opposite the field direction, which
suggests a negative effective chain charge.  We found that the stronger the
field, the faster the chain will drift.  For a long chain, $\mu_{pe}$ shows
furthermore a plateau region when $E>E^*$.  The dependence of
$\mu_{pe}$ on the electric field and the chain length gives a plausible way to
electrophoretically separate PE chains by size in free solutions by means
of chain unfolding transition~\cite{hsiao08}. 

\begin{figure} \centering 
\includegraphics[width=0.6\textwidth,angle=270]{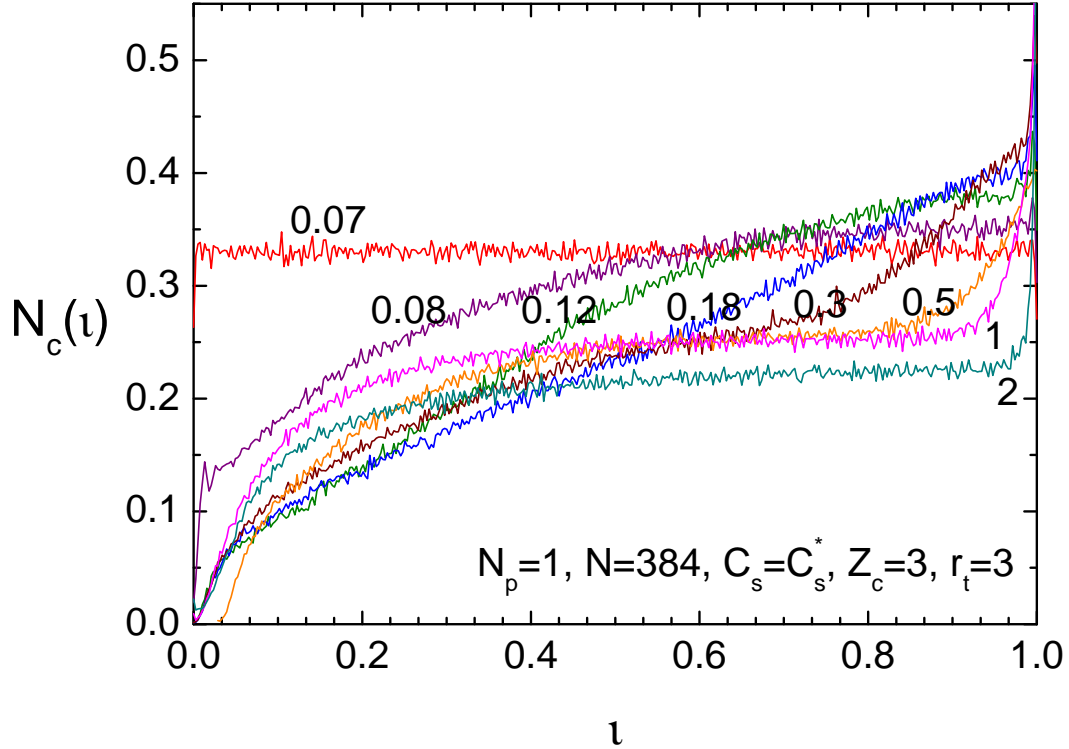}
\caption{$N_c(\iota)$ obtained in different field strength.
The value of $E$ is indicated near each curve.} \label{f:N3-Ipos} \end{figure}

The variation of $\mu_{pe}$ can be related to the ion condensation on the chain
under the action of the electric field.   Therefore, we studied here the number
of the condensed trivalent ions on the chain by counting the ions inside the
worm-shaped tube of radius $r_t=3$ around the chain. The results for $N=384$
in different strength of electric field are plotted in Fig.~\ref{f:N3-Ipos}
against the monomer index $\iota$, rescaled from 0 to 1, where $\iota=0$
denotes the first monomer heading toward the field direction and $\iota=1$
denotes the last monomer of the other chain end.  

We saw that $N_c(\iota)$ is flat when the applied field is small, $E\le 0.007$,
which shows an uniform distribution of the condensed trivalent ions along the
chain.  There is about 0.33 trivalent counterions condensed on each monomer,
which indicates the neutralization of the negatively-charged chain backbone 
by these condensed counterions.  If we further increases the
electric field, these condensed ions distribute non-uniformly on the chain
where fewer ions condensed near the heading end ($\iota=0$)  than the tailing
end ($\iota=1$).  When $0.07<E<0.2$, $N_c(\iota)$ looks similar to an inclined
line and the slope increases with $E$, resulting in a decrease of the total
number of the condensed trivalent counterions on the chain.  Therefore,
$\vert\mu_{pe}\vert $ increases with $E$ due to this partial detachment of the
condensed ions by the electric field. At this moment, the PE-ion complex is
polarized in a way that the condensed trivalent counterions are bound,
basically immobile, on the chain.  For the higher electric field,
$0.2<E<1.0$, $N_c(\iota)$ becomes a horizontal sigmoidal curve and the value in
the middle chain region is  independent of $E$.  The appearance of
this horizontal region reflects the fact that the condensed trivalent
counterions are now gliding on the chain. These ions can be stripped off the
chain by the strong electric field and the other ions in the bulk solution then
condense onto it, establishing a steady state.  The total number of the
condensed trivalent counterions is approximately a constant in this electric
field, which results in the plateau region of $\mu_{pe}$ against $E$.  For an
even stronger electric field, such as $E=2.0$, the baseline of the horizontal
sigmoidal curve moves downward. The condensed trivalent ions are stripped off
the chain even more.  The effective chain charge is thus more negative and
$\vert \mu_{pe}\vert$  increases.  

\subsection{Conformation of an unfolded PE chain}

\begin{figure} \centering 
\includegraphics[width=0.7\textwidth,angle=270]{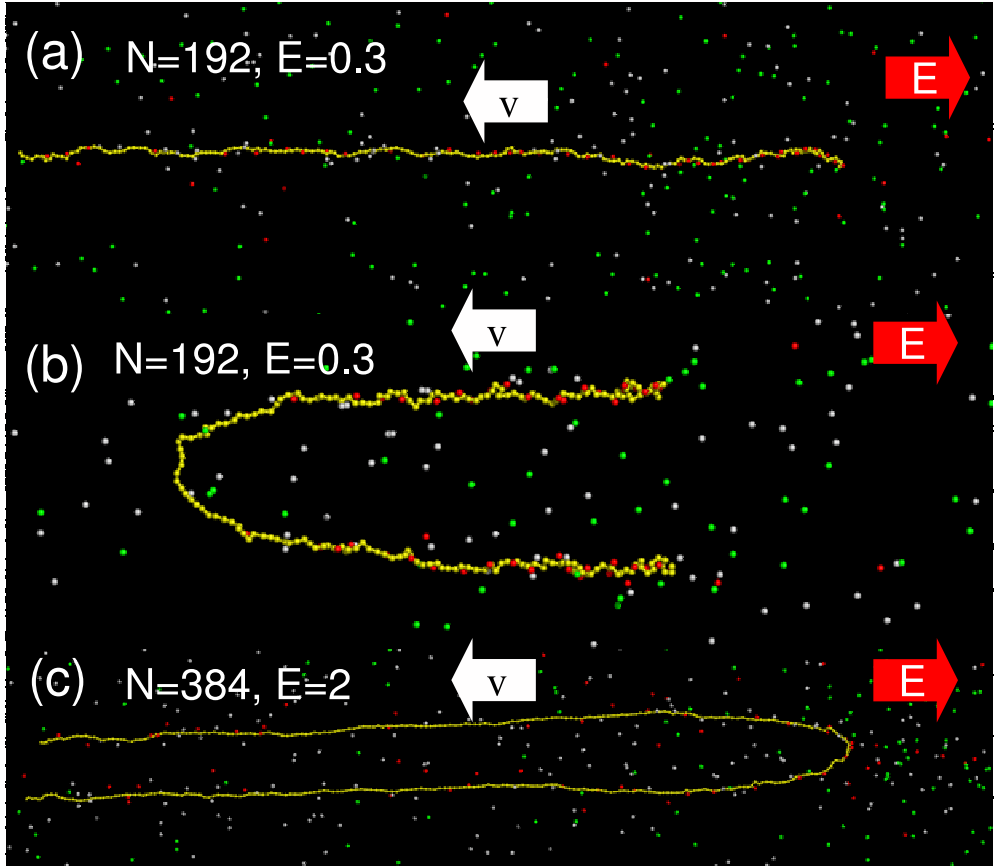}
\caption{(Color on line) Snapshots of unfolded PE chains in electric fields.
The yellow, the white, the red, and the green spheres represent, respectively, the monomers,
the monovalent counterions, the trivalent counterions, and the coins.
The chain length, the electric field, the field direction, and the chain drifting 
direction are indicated in the figure.}
\label{f:snapshots} \end{figure}

In our simulations, the PE chains were unfolded, for the most of the time, to
an extended structure, similar to a straight line, aligned parallel to the
field direction (see in Fig.~\ref{f:snapshots}(a)).  Nonetheless, we observed
sometimes that they were unfolded to a U-shaped structure in the electric
fields.  The open side of the U shape can point opposite or toward the chain
drifting direction as shown in Fig.~\ref{f:snapshots}, panel (b) and (c),
respectively.

\begin{figure} \centering 
\includegraphics[width=0.6\textwidth,angle=270]{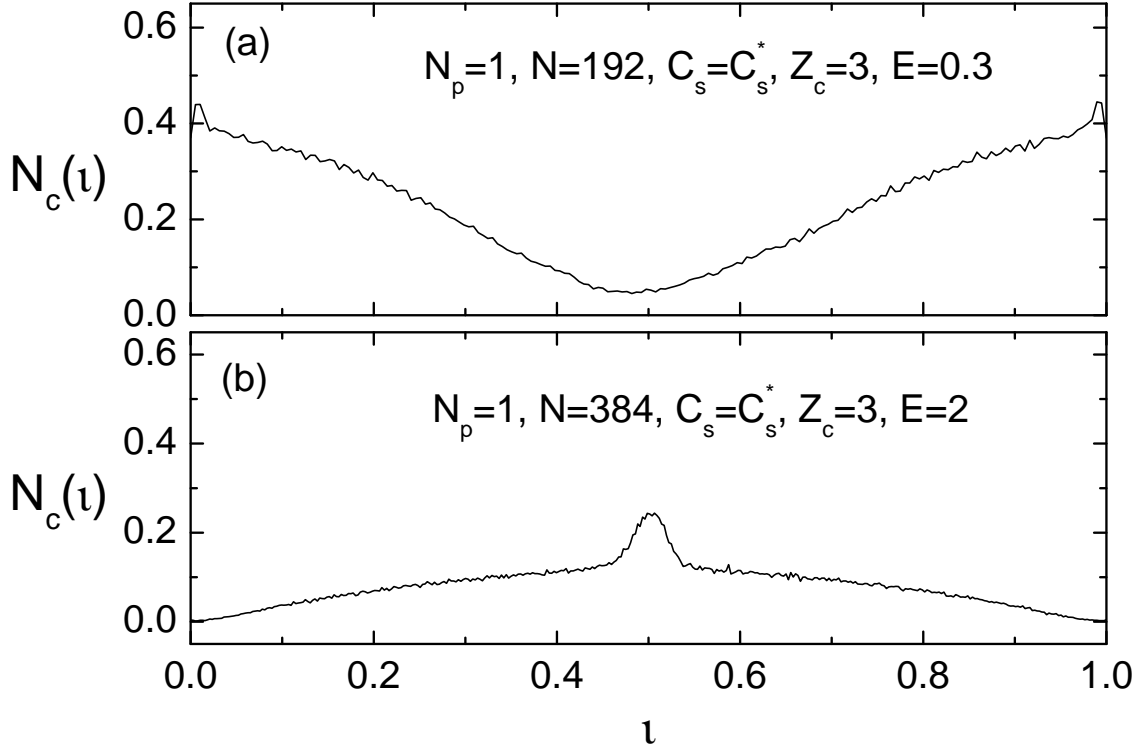}
\caption{$N_c(\iota)$ for the two U-shaped chains in
Fig.~\ref{f:snapshots}(b) and (c), respectively.} \label{f:N3-Ipos_U}
\end{figure}

This U-shaped structure has been observed experimentally in electrophoresis of
microtubules~\cite{heuvel08} and also been shown in simulations of the elastic
uncharged/charged chains in stokes flows or in electric
fields~\cite{lagomarsino05,schlagberger05,schlagberger08,frank08}.  These
studies showed that a combination of the elastic  and the hydrodynamic effect
results in the bending of a rigid chain into a horseshoe shape, oriented
perpendicular to the direction of motion~\cite{schlagberger05,lagomarsino05}.
If chains are charged and the driving force is an electric field, other effect,
the electric polarization of the PE complex, will play a role, which favors
parallel orientation to the electric field, and compete with the
elasto-hydrodynamic effect~\cite{schlagberger08}.  In our simulations, the PE
chains are flexible and the hydrodynamic interaction is neglected.  Therefore,
different mechanism drives the chains to form U-shaped structures
where the field-induced dipole moments on the two branches of a 
U-shaped chain establish an equilibrium. This phenomenon can be seen by plotting in
Fig.~\ref{f:N3-Ipos_U} the distribution of the condensed trivalent counterions
$N_c(\iota)$ for the two U-shaped chains from Fig.~\ref{f:snapshots}(b) and
(c), respectively.

The symmetry of $N_c(\iota)$ with respect to the middle point of the chain
($\iota=0.5$) shows that an equilibrated polarization was established on the
two branches of the U-chain in the electric field.   The existence of two
pointing directions of the open side of the U chain is a feature specially for
the electric polarization. It is distinguishable to the elasto-hydrodynamic
effect where only the U-shaped structure with the open side opposite to the
moving direction is produced.  Moreover, we notice that the electrophoretic
mobility of an U-shaped chain is approximately equal to that of an elongated
chain of half of the chain length.  For example, $\mu_{pe}$ is $-0.209(4)$ in
Fig.~\ref{f:snapshots}(b), close to the mobility of the elongated chain of
$N=96$, $-0.225(3)$.  Furthermore, we verified the stability of these U-shaped
chains and found that they can persist through the whole simulation period
corresponding to, at least, the order of microsecond. However, by introducing
some perturbations such as AC electric fields, the U-shaped structure can be
transfered into the elongated chain structure but the inverted direction of
transfer cannot be realized. Therefore, the U-shaped structure is probably
metastable.  We have calculated the total energy of the system for the U-shaped
chain structure and also for the extended-chain structure.  We found that the
previous energy is, at least, $5\%$ higher than the latter.  Moreover, the
U-shaped chain has a slower electrophoretic mobility than the extended chain,
which implies a larger number of counterions condensed on the U-shaped chain
to decrease the effective chain charge; consequently, fewer ions are presented
in the bulk solution and the entropy of the solution is small, compared to the
extended-chain structure.  Therefore, the free energy of the
system is lower for the extended chain than for the U-shaped chain.  
This estimation supports that the U-shaped structure is metastable.
Since the open side of the U-shaped chain can point to one of the two directions,
along or against the field direction, we predict the existence of other metastable
states, due to polarization, in which the chain shows many bends, such as
S-shaped or W-shaped structures, in electric fields.   

\section{Conclusions}

We have studied the behavior of single polyelectrolytes condensed by trivalent
salt under the action of an uniform electric field by means of Langevin
dynamics simulations.  We found that the chains unfolded while the strength of
the electric field is stronger than some critical value $E^*$, similar to the
previous study where the chains were condensed by tetravalent
salt~\cite{hsiao08}.  $E^*$ shows scaling-law dependence on the chain length
$N$, reading as $E^* \sim N^{-0.77(1)}$.  The exponent in the scaling law is
different from the prediction by Netz~\cite{netz03a,netz03b} and from the
simulations in tetravalent salt solutions~\cite{hsiao08}, which demonstrated
the importance of the salt valence on the exponent.  Therefore, the weaker
the condensing agent, the larger the absolute value of the exponent and the
easier the unfolding of a  condensed chain will be.  We showed that the
electrophoretic mobility of chain $\vert\mu_{pe}\vert$ drastically increases
while the chain is unfolded.  The distribution of the condensed counterions on
the chain was studied and related to the change of the mobility in different
regions of electric field.  The dependence of $\mu_{pe}$ on the chain length
and the electric field enables us to device a way to impart chain-length
dependence in free-solution electrophoresis through chain-unfolding
mechanism in electric fields. Finally, we pointed out the possibility to unfold
a condensed PE chain into an U-shaped structure in electric fields, in addition
to the elongated structure, with the open side of the U heading or tailing the
chain drifting direction.  This structure is a result of purely electric
polarization, different from the formation of the horseshoe-shaped chains in
sedimentation experiments caused by elasto-hydrodynamics. 

\section{Acknowledgments}

This material is based upon work supported by the National Science Council, the
Republic of China, under the contract No.~NSC 97-2112-M-007-007-MY3.  The
computing resources are supported by the National Center for High-performance
Computing. 


\end{document}